\documentclass[twocolumn,tighten]{aastex631}

\usepackage{enumerate}

\usepackage{amsmath}

\shorttitle{Equilibrium Temperature of Planets}
\shortauthors{Andreas Quirrenbach}

\graphicspath{{./}{figures/}}

\begin{document}

\title{The Equilibrium Temperature of Planets on Eccentric Orbits: Time Scales and Averages}


\author{Andreas Quirrenbach}
\affiliation{Landessternwarte \\
Zentrum f\"ur Astronomie der Universit\"at Heidelberg \\
K\"onigstuhl 12 \\
D-69117 Heidelberg, Germany}

\begin{abstract}

From estimates of the near-surface heat capacity of planets it is shown that the thermal time scale is larger than the orbital period in the presence of a global ocean that is well-mixed to a depth of 100\,m, or of an atmosphere with a pressure of several tens of bars. As a consequence, the temperature fluctuations of such planets on eccentric orbits are damped. The average temperature should be calculated by taking the temporal mean of the irradiation over an orbit, which increases with $1/\sqrt{1-e^2}$. This conclusion is independent of the orbital distance and valid for Sun-like stars; the damping is even stronger for low-mass main sequence hosts.

\end{abstract}

\section{Introduction} \label{sec:intro}

The assumption that planets are in radiative equilibrium with their host stars provides a first estimate of their temperature, and such calculations are customarily included in most exoplanet discovery papers. The textbook equation for the equilibrium temperature is
\begin{equation}\label{equ}
T_{\rm eq} = \left[ \frac{Q}{\sigma} \right]^{1/4} = \left[ \frac{(1-A) R_\ast^2}{4r^2} \right]^{1/4} \cdot T_\ast~~,
\end{equation}
where $Q$ is the absorbed radiation per unit area, $\sigma$ the Stefan-Boltzmann constant, $A$ is the Bond albedo of the planet, $R_\ast$ and $T_\ast$ are the radius and temperature of the star, and $r$ the distance of the planet from the star. This equation assumes a blackbody spectrum, and it does not correspond directly to surface temperature as it does not account for day- versus nightside temperature differences, internal heat sources, or greenhouse warming. These issues will not be considered here as the discussion will concentrate on the effect of a non-zero orbital eccentricity.

\section{Expressions for the Average Temperature}
\label{sec:equi}

In the case of a circular orbit, the distance $r$ is simply equal to the semi-major axis, and the equilibrium temperature is given by
\begin{equation}\label{circular}
T_{\rm c} = \left[ \frac{(1-A) R_\ast^2}{4a^2} \right]^{1/4} \cdot T_\ast~~.
\end{equation}
The temperature of a planet with vanishing heat capacity and instantaneous heat distribution across its surface, on an orbit with eccentricity $e$, would vary between the periastron and apastron values
\begin{equation}\label{periastron}
T_{\rm p} = \frac{1}{\sqrt{1-e}}  \cdot T_{\rm c}~~,~~T_{\rm a} = \frac{1}{\sqrt{1+e}}  \cdot T_{\rm c}~~.
\end{equation}
Climate models and studies of the impact of orbital eccentricity on habitability \citep[e.g.][]{Williams2002,Barnes2008,Bolmont2016} have usually been based on the mean flux received by the planet, averaged over the orbit. As this quantity is $\propto \left( 1-e^2 \right)^{-1/2}$ \citep{Johnson1976}, the corresponding ``flux-averaged'' temperature is
\begin{equation}\label{fluxav}
\begin{split}
T_{\rm f} & = \left( 1-e^2 \right)^{-1/8} \cdot T_{\rm c} \\
          & = \left[ 1 + \frac{1}{8} e^2 + \frac{9}{128} e^4 + \mathcal{O}(e^6) \right]  \cdot T_{\rm c}~~.
\end{split}
\end{equation}
More recently, \citet{Mendez2017} have computed the time average of Eqn.~\ref{equ},
\begin{equation}\label{tempav}
\begin{split}
T_{\rm t} & = \frac{2\sqrt{1+e}}{\pi} \, \mathcal{E} \left( \sqrt{\frac{2e}{1+e}} \right) \cdot T_{\rm c} \\
          & = \left[ 1 - \frac{1}{16} e^2 - \frac{15}{1024} e^4 + \mathcal{O}(e^6) \right]  \cdot T_{\rm c}~~,
\end{split}
\end{equation}
where $\mathcal{E}(k)$ is the complete elliptic integral of the second kind.\footnote{Note that sometimes the notation $\mathcal{E}(m), m=k^2$ is used.} They point out that this expression is a decreasing function of $e$, and assert that it should be used instead of Eqn.~\ref{fluxav}, which has the opposite behavior. This suggestion has been followed by a number of authors \citep[e.g.][]{Brahm2019,Schlecker2020,Hobson2021,Schanche2022}, but in most cases this is not appropriate because the ansatz inherent in Eqn.~\ref{tempav} neglects the thermal inertia of the planet.

\section{Thermal and Orbital Time Scales}
\label{sec:scales}

Consider first the two limiting cases where the thermal time scale $\tau$ is (i) very short or (ii) very long compared to the orbital period $P$:
\begin{enumerate}[(i)]
  \item The planetary temperature adjusts instantaneously, and oscillates between the periastron and apastron values (Eqn.~\ref{periastron}). Calculating the orbital average from Eqn.~\ref{tempav} is thus merely a numerical exercise without much physical meaning.
  \item The amplitude of the temperature oscillations is strongly reduced compared to case (i). Using an orbital average is thus much more meaningful, but it should be computed from Eqn.~\ref{fluxav}.
\end{enumerate}
To estimate $\tau$, one can introduce the heat capacity per unit area $C$, linearize the energy balance equation,
\begin{equation}\label{balance}
C\,\frac{{\rm d}T}{{\rm d}t} = Q - \sigma T^4~~,
\end{equation}
around the equilibrium value $T_0$ and obtain
\begin{equation}\label{thermal}
\tau = \frac{CT_0}{4Q}~~.
\end{equation}
If a small forcing term is added as a step function, the system will relax exponentially with an e-folding time $\tau$ to the new equilibrium temperature.

For a planet on an eccentric orbit, $\tau$ is time scale on which the planetary temperature adjusts to the changing irradiation, and the ratio $\tau/P$ describes the transition between the regimes (i) and (ii).

Noting that $\tau \propto Q^{-3/4} \propto a^{3/2}$ and keeping Kepler's third law in mind, one sees that coincidentally $\tau/P$ does not depend on $a$. For any given host star, $C$ is thus the sole parameter defining the boundary between cases (i) and (ii). The mass-luminosity relation on the main sequence can reasonably be represented by a power law of the form $L_\ast \propto M_\ast^\alpha$ with $\alpha$ in the range 3.5--4. Adopting the value $\alpha \approx 3.66$ from \citet{Andrade2019}, 
one finds
\begin{equation}\label{ratio}
\tau/P \propto M_\ast^{-3\alpha/4 + 1/2} \approx M_\ast^{-2.25}~~,
\end{equation}
i.e., ``identical'' planets orbiting lower-mass stars tend to be closer to limit (ii). This is to be expected, of course, and corresponds to the well-known fact that the orbital period corresponding to a given equilibrium temperature decreases with stellar mass.

\section{Examples in the Solar System}
\label{sec:sol}

The heat capacity $C$ depends strongly on the heat transport close to the surface. As a concrete example, consider a planet with a global ocean that is well-mixed to a depth of 100\,m. From the specific heat capacity of water, $c_{\rm H_2O} = 4.2$\,J\,g$^{-1}$\,K$^{-1}$, one gets $C = 4.2 \cdot 10^8$\,J\,K$^{-1}$\,m$^{-2}$. Inserting this and the terrestrial values $T = 255$\,K and $Q = 237$\,W\,m$^{-2}$ in Eqn.~\ref{thermal}, one gets $\tau = 1300$\,d, in good agreement with the empirically determined value $\tau = 5 \pm 1$\, yr \citep{Schwartz2007}.

Alternatively, the heat capacity may be dominated by a dense atmosphere. It can be shown easily from the barometric height formula that in this case
\begin{equation}\label{atmos}
C = \frac{c_{\rm P}\,p}{g}~~,
\end{equation}
where $c_{\rm P}$ is the specific heat (per mass) at constant pressure, $p$ the pressure at the bottom of the well-mixed atmospheric layers, and $g$ the surface gravity. With $c_{\rm P, air} = 1$\,J\,g$^{-1}$\,K$^{-1}$ and $p = 1$\,bar, one obtains $C_{\rm atm} = 1 \cdot 10^7$\,J\,K$^{-1}$\,m$^{-2}$ for the Earth's atmosphere, which is much smaller than the heat capacity of the oceanic surface layers. For Venus, however, the surface pressure and thus the atmospheric heat capacity is two orders of magnitude higher ($c_{\rm P, CO_2} = 0.85$\,J\,g$^{-1}$\,K$^{-1}$, $p = 93$\,bar, $g = 8.9$\,m\,s$^{-2}$, $C = 8.9 \cdot 10^8$\,J\,K$^{-1}$\,m$^{-2}$).

The convection cells on Jupiter extend to a depth of at least 240\,bar \citep{Duer2021}. Considering also the high specific heat capacity of its hydrogen-dominated atmosphere, this means that Jupiter has the highest value of $\tau / P$ of the planets considered here.

\section{Conclusions}
\label{sec:concl}

The above rough estimates strongly suggest that substantial liquid surface water reservoirs or atmospheres of several tens of bars can effectively buffer variations of the stellar irradiation on orbital time scales. Temperature variations of planets with such characteristics on eccentric orbits should thus be damped, with the effective temperature fluctuating around the flux-averaged mean (Eqn.~\ref{fluxav}). This conclusion is independent of the semi-major axis. It is valid for Sun-like stars, with even stronger damping for low-mass stars (Eqn.~\ref{ratio}).

In light of this finding, the statement by \citet{Mendez2021}, ``when orbital eccentricity increases, the average equilibrium temperature decreases, thus extending the size of the Habitable Zone'', appears incorrect, as habitability is certainly favored by the presence of climate buffers. It should be noted, however, that only sophisticated climate models can address the much more complicated questions arising from longitudinal and latitudinal temperature differences, which also depend on the ob\-li\-qui\-ty and on the distribution of continents and oceans, as in the Milankovitch cycles of the Earth's climate \citep[e.g.,][]{Berger1988}. Detailed investigations of planetary climates that include a periodic forcing due to orbital eccentricity, with tools such as VPLanet \citep{Barnes2020} or with idealized global circulation models \citep{Guendelman2020}, will thus be a rewarding undertaking.

\begin{acknowledgments}
{\it I thank Rory Barnes and Karan Molaverdikhani for helpful comments on the manuscript.}
\end{acknowledgments}



\bibliography{Equilibrium_Temperature}{}
\bibliographystyle{aasjournal}


\end{document}